\documentclass{article}

\usepackage{amsmath, amsthm, amssymb}
\usepackage{url}
\usepackage{graphicx}
\usepackage{psfrag}
\usepackage{color}

\newtheorem{example}{Example}
\newtheorem{theorem}{Theorem}
\newtheorem{lemma}{Lemma}

\newtheorem*{cond:a}{Condition A}
\newtheorem*{cond:b}{Condition B}
\newtheorem*{cond:t1}{Condition T1}
\newtheorem*{cond:t2}{Condition T2}

\renewcommand{\qed}{$\Box$}
\renewenvironment{proof}{\par\noindent{\it Proof.}}{\hfill\qed\medskip}

\title{Truth-telling Reservations}
\author{Fang Wu \quad Li Zhang \quad Bernardo A. Huberman\\
HP Labs, Palo Alto, CA 94304}
\begin{document}
\maketitle

\begin{abstract}
We present a mechanism for reservations of bursty resources that
is both truthful and robust. It consists of option contracts whose
pricing structure induces users to reveal the true likelihoods
that they will purchase a given resource. Users are also allowed
to adjust their options as their likelihood changes. This scheme
helps users save cost and the providers to plan ahead so as to
reduce the risk of under-utilization and overbooking. The
mechanism extracts revenue similar to that of a monopoly provider
practicing temporal pricing discrimination with a user population
whose preference distribution is known in advance.
\end{abstract}

\pagebreak
\section{Introduction}\label{sec:intro}

A number of compute intensive applications often suffer from
bursty usage patterns~\cite{beran-95, clearwater-05, gribble-98,
huberman-97, voldman-81},
whereby the demand for IT, memory and bandwidth resources
can at times exceed the installed capacity within the
organization. This problem can be addressed by providers of IT
services who satisfy this peak demand for a given price, playing a
role similar to utilities such as electricity or natural gas.

The emergence of a utility form of IT provisioning creates in turn
a number of problems for both providers and customers due to the
uncertain nature of IT usage. On the provider side there is a need
to design appropriate pricing schemes to encourage the use of such
IT services and to gain better estimates of the usage pattern so
as to enable effective statistical multiplexing. On the customer
side, there needs to be a simple way of figuring out how to
anticipate and hedge the need for uncertain demand as well as the
costs that it will add to the overall IT operations.

Recently, it was proposed to use swing options for pricing the
reservation of IT resources \cite{clearwater-05}.  By purchasing a
swing option the user pays an upfront premium to acquire the right,
but not the obligation, to use a resource as defined in the option
contract. As with the case with electricity, IT resources, such as
bandwidth and CPU time, are non-storable and with volatile usage
pattern. Thus, swing options provide flexibility in both the amount
and the time period for which a resource is purchased, making them
appealing to users whose bursty demand is hard to predict. From the
point of view of the providers, if enough users purchase these
options providers can offset the cost of providing peak capacity by
multiplexing among many users.

Pricing a swing option for IT resources however, turns out to be
difficult because of the complexity of the option contract and the
lack of a good model of the spot market price, which at present is
nonexistent. Moreover, there are two important problems that need
resolution. First, the user needs to be able to estimate the
amount of resources that need to be reserved as well as their
cost; and second,  the provider needs to put in place a mechanism
that will induce truth revelation on the part of the user when
stating the likelihood that a given reservation or option will be
exercised.

As was shown in \cite{clearwater-05}, the first problem can be
addressed by providing the user with a simulation tool for
estimating the cost of a reservation from a set of historical
data, as well as a provision for entering the user's assumptions
about aggressive or conservative swings. Because the prices for
swings are set ahead of time and not by market forces, the
forecasting tool also provides a powerful ``what-if'' capability
to both the resource provider and the customer for estimating
outright costs and risks associated with fluctuations in customer
demand.

As to the provider's problem with asymmetric information, it could
be argued that a user's historical usage pattern allows to predict
his future demand. But in many cases, such as with new users, the
data may not be available or reflect unanticipated user needs.
Even worse, users may misrepresent the likelihood of their needs
in order to gain a pricing advantage, with the consequent loss to
the provider. In the original design of the swing option this was
addressed by introducing a time dependent discount that induces
early commitment to a contract. But this strategy still allows
users to misrepresent their likelihoods the first time they buy an
option.

This paper presents a solution to the truth revelation problem in
reservations by designing option contracts with a pricing
structure that induces users to reveal their true likelihoods that
they will purchase a given resource. A user is allowed to adjust
his option later if his likelihood changes. Truthful revelation
helps the provider to plan ahead to reduce the risk of
under-utilization and overbooking, and also helps the users to
save cost.  In addition to its truthfulness and robustness, the
mechanism extracts revenue similar to that of a monopoly provider
monopoly provider practicing temporal pricing discrimination with
a user population whose preference distribution is known in
advance \cite{conlisk-84, gale-93, gribble-98, sobel-91,
varian-89, voldman-81, wilson-93}.

We  start by presenting a simple two period model in which at the
first period the user knows his probability of using the resource
at the second period and purchases a reservation whose price
depends on that probability. A coordinator then aggregates the
reservations from all users and purchases the needed resources
from the provider. These resources are then purchased in the
second period. A nonlinear pricing scheme is shown to lead to both
truthful revelation and profitability to both users and
coordinator.

We then extend the two period model to a multi period model which
allows the user's likelihoods of use to change over time. In this
more realistic scheme, users are allowed adjust their options
according to updated information about their needs while remaining
truthful at each time period.

Finally we show how this truth-telling reservation mechanism can
be interpreted in terms of standard options terminology, and
finish with a discussion of the feasibility of this mechanism to
provide revenues to both users and providers.

While the focus of this paper is on reservations for IT resources,
there are many other interesting situations where our mechanism can be
useful. For example, conference rooms in many organizations tend
to be reserved in advance in the likelihood that they will be
needed for future meetings. The resulting behavior leads to
serious inefficiencies through reservations that are not exercised
and which force others to reschedule important meetings. A truth
telling mechanism like the one we propose would lead to a more
efficient scheduling of such conferences. Likewise, airline seats,
hotel reservations, network bandwidth and tickets for popular
shows would benefit from a properly priced reservation system,
leading to both more predictable use and revenue generation.

\section{The Two Period Model}

Consider $n$ users $\{1, 2, \dots, n\}$ who live for two discrete
periods. Each user may need have to consume one unit of resource
in period 2, which he can buy from a resource provider either in
period 1 at a discount price $1$, or in period 2 at a higher price
$C>1$. In period 1, each user $i$ only knows the probability $p_i$
that he will need the resource in the next period. It is not until
period 2 that he can be certain about his need (unless $p=0$ or
$1$). We also assume that the distributions of the users' needs
are independent.

Suppose all the users wish to pay the least while behaving in a
risk-neutral fashion. User $i$ can either pay $1$ in period 1, or
wait until period 2 and pay $C$ if it turns out he has to, an
event that happens with probability $p_i$. Obviously, he will use
the former strategy when $Cp_i>1$ and the latter strategy when
$Cp_i<1$, while his cost is $\min(1, Cp_i)$.

This optimal paying plan can be very costly for the user. For
example, when $C=5$ and $p=0.1$, the user always postpones the decision
to buy
until period 2, ending up paying $5$ for every unit he needs.

In what follows we describe a reservation mechanism that allows
him to pay a small premium that guarantees his one unit of
resource whenever he needs it in period 2, at a price not much
higher than the discount price $1$.  In addition, the mechanism
makes the user truthfully reveal his probability of using the
resource to the provider, who can then accurately anticipate user
demand. At a later stage, we show how this mechanism can be
thought of as an option.

\subsection{\label{sec:2period}The coordinator game}

To better illustrate the benefit of this mechanism, we introduce a
third agent, the coordinator, who aggregates the users'
probabilities and makes a profit while absorbing the users' risk.
He does so in a two period game, which we now describe.

\begin{enumerate}
 \item (Period 1) The coordinator asks each user to submit a
 probability $q_i$, which does not have to be the real
 probability $p_i$ that the user will need one unit of resource in
 period 2.

 \item (Period 1) The coordinator reserves $\sum q_i$ units of
 resource from the resource provider (at the discount price), ready to
 be consumed in period 2.

 \item (Period 2) The coordinator delivers the reserved resource units
 to users who  claim them. If the amount he reserved is not
 enough to satisfy the demand, he buys more resource from the provider
(at the higher unit
 price $C$) to meet the demand.

 \item (Period 2) User $i$ pays
\begin{equation}
\begin{cases}
f(q_i) & \text{if he needs one unit of resource,}\\ g(q_i) & \text{if
he does not need it,}
\end{cases}
\end{equation}
where $f,g: [0,1]\to \mathbb R^+$ are two functions whose forms will
be specified later.
\end{enumerate}
These terms are completely transparent to everyone, before step 1.

For the coordinator to profit, the following two conditions have to be
satisfied:
\begin{cond:a}
The coordinator can make a profit by providing this service.
\end{cond:a}
\begin{cond:b}
Each user prefers to use the service provided by the coordinator,
rather than to deal with the resource provider directly.
\end{cond:b}
The next two truth-telling conditions, although not absolutely
necessary, are useful for conditions A and B to hold.
\begin{cond:t1}
(Step 1 truth-telling) Each user submits his true probability $p_i$ in
step 1, so that he expects to pay the least later in step 4.
\end{cond:t1}
\begin{cond:t2}
(Step 3 truth-telling) In step 3, when a user does not need a
resource in period 2, he reports it to the coordinator.
\end{cond:t2}

From Condition T1. User $i$ expects to pay
\begin{equation}
w(q_i) \equiv p_i f(q_i) + (1-p_i) g(q_i)
\end{equation}
in period 2. His optimal submission $q_i^*$ is determined by the
first-order condition
\begin{equation}
w'(q_i^*) = p_i f'(q_i^*) + (1-p_i) g'(q_i^*)=0.
\end{equation}
Truth-telling requires that $q_i^* = p_i$, or
\begin{equation}
p_i f'(p_i) + (1-p_i) g'(p_i) = 0.
\end{equation}

Condition T2 simply requires that
\begin{equation}
f(p) \ge g(p) \quad \text{for all } p\in[0,1].
\end{equation}

Now we study Condition A when all users submit their true
probabilities $\{p_i\}$. Let $U$ be the total resource usage of all
users in period 2, and let $W$ be the their total payment. Both $U$
and $W$ are random variables. Clearly,
\begin{equation}
\mathbb E\,U = \sum p_i,
\end{equation}
and
\begin{equation}
\mathbb E\,W = \sum w(p_i).
\end{equation}

\begin{lemma}
\label{lm:conda}
If there exists an arbitrarily small $\epsilon>0$
such that
\begin{equation}
\label{eq:conda-strong} w(p) \ge p+\epsilon \quad \text{for all }
p\in[0,1],
\end{equation}
then
\begin{equation}
W-U \to \infty \text{ a.s. \quad as } n\to \infty,
\end{equation}
i.e.~by charging an arbitrarily small premium, the coordinator makes
profit when there are many users (Condition A).
\end{lemma}
\begin{proof}
This follows directly from the ``$X^4$-strong law''. (See e.g.~David
Willams, Probability with Martingales, pp.~72--73, Cambridge
University Press, 2001. The random usage of each user does not have
to be identically distributed.)
\end{proof}

\medskip
The small number $\epsilon$ is merely a technical device. In what
follows we will neglect it and use a weakened version of
Eq.~(\ref{eq:conda-strong}) as a sufficient condition of Condition A
(not rigorous):
\begin{equation}
w(p) \ge p \quad \text{for all } p\in[0,1].
\end{equation}

Last, Condition B says that the user can save money by using the
coordinator's service:
\begin{equation}
w(p) \le \min(1,Cp) \quad \text{for all } p\in[0,1].
\end{equation}

To summarize, the following conditions on $f$ and $g$ are
sufficient for the truth-telling mechanism to work:
\begin{align}
& \label{eq:condt1} p f'(p) + (1-p) g'(p) = 0,\\
& \label{eq:condt2} f(p) \ge g(p),\\
& \label{eq:condab} p \le p f(p) + (1-p) g(p) \le \min(1,Cp),
\end{align}
for all $p\in [0,1]$.

Consider the following choice\footnote{This choice is not unique,
but is analytically simple.}:
\begin{equation}
\label{eq:diff-eqns} f'(p) = -k(1-p), \quad g'(p) = kp,
\end{equation}
which obviously satisfies Eq.~(\ref{eq:condt1}). Letting $p=0$ and 1 in
Eq.~(\ref{eq:condab}) gives two boundary conditions for $f$ and $g$:
\begin{equation}
\label{eq:bc} f(1)=1, \quad g(0)=0.
\end{equation}
The solution for Eq.~(\ref{eq:diff-eqns}) and (\ref{eq:bc}) is
\begin{eqnarray}
\label{eq:f} f(p) &=& 1 + \frac k2 -kp + \frac {kp^2}2,\\
\label{eq:g} g(p) &=& \frac {kp^2} 2.
\end{eqnarray}
To check Eq.~(\ref{eq:condt2}) and (\ref{eq:condab}), we first calculate
\begin{equation}
w(p) = \left( 1+\frac k2 \right) p - \frac k2 p^2.
\end{equation}
And then it is not hard to show

\begin{lemma}
For the choice of $f$ and $g$ in Eq.~(\ref{eq:f}) and (\ref{eq:g}),
conditions (\ref{eq:condt2}) and (\ref{eq:condab}) are satisfied for
$k\in [1, \min\{2(C-1),2\}]$.
\end{lemma}
\begin{proof}
Eq.~(\ref{eq:condt2}) is satisfied because
\begin{equation}
f(p)-g(p) = 1+k \left(\frac 12-p\right) \ge 1-\frac k2 \ge 0.
\end{equation}
To verify Eq.~(\ref{eq:condab}), we write
\begin{eqnarray}
w(p)&=&p+\frac k2 p(1-p) \ge p,\\
w(p)&=&1-(1-p)\left(1-\frac k2 p\right) \le 1,\\
w(p)&\le&Cp-\frac k2 p^2 \le Cp.
\end{eqnarray}
\end{proof}

\begin{figure}
\psfrag{p}{\textcolor{blue}{$p$}}
\psfrag{w}{\textcolor{red}{$w(p)$}}
\psfrag{min}{\textcolor{blue}{$\min(1, Cp)$}}
\psfrag{pp}{$p$}
\centering \includegraphics[scale=.7]{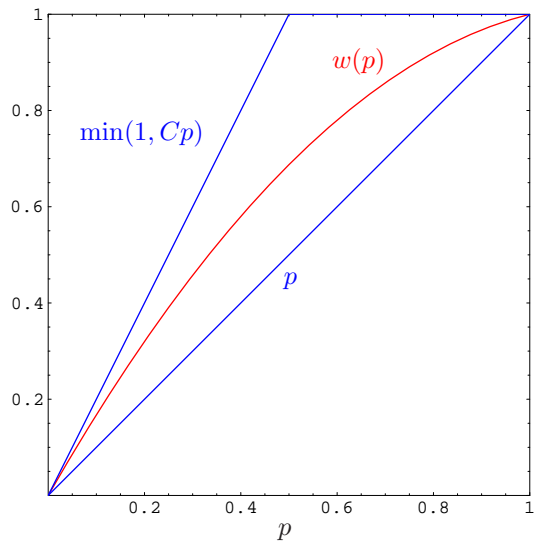}
\caption{\label{fig:w} Plot of $w(p)$ for $C=2$ and $k=1.5$. The red
curve $w(p)$ lies completely in the region enclosed by the two blue
curves.}
\end{figure}

\medskip

Fig.~\ref{fig:w} shows the special case $C=2$ and $k=1.5$. As can
be seen, the red curve $w(p)$ lies completely between the two blue
curves. The difference between the upper blue curve and the red
curve is the amount of money the user saves (varying with
different $p$). The difference between the red curve and the lower
blue line is the coordinator's expected payoff from one user. Note
that his payoff is larger for values of $p$'s lying in the middle
of the range, and is zero for $p=0$ and $p=1$. This result is
hardly surprising, for when there is no uncertainty the user does
not need a coordinator at all. Thus, the coordinator makes a
profit out of uncertainties in user behavior.

\begin{figure}
\psfrag{p}{$p$}
\psfrag{f}{\textcolor{red}{$f(p)$}}
\psfrag{g}{\textcolor{blue}{$g(p)$}}
\centering \includegraphics[scale=.7]{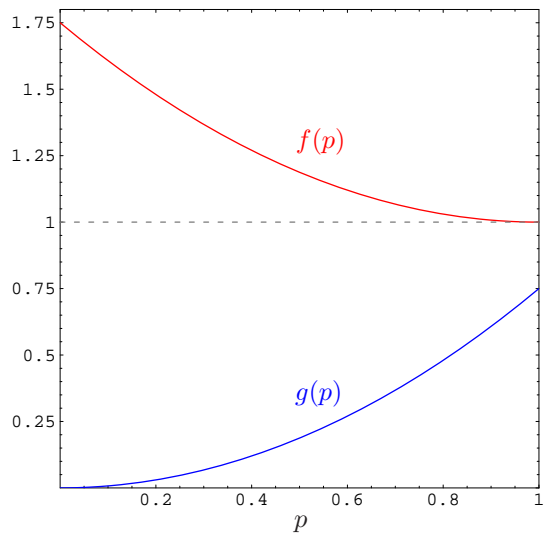}
\caption{\label{fig:fg}Payment curves of the user for $C=2$ and
 $k=1.5$. If the user needs one unit of resource in period 2, he pays
 according to the upper curve. Otherwise he pays according to the
 lower curve.}
\end{figure}

Fig.~\ref{fig:fg} plots the two payment curves, $f(p)$ and $g(p)$,
for $C=2$ and $k=1.5$. After signing a contract, a user agrees to
pay later either the upper curve for one unit of resource, or the
lower curve for nothing. Note that $f(p)$ is strictly decreasing,
a feature essential for the user to be truth-telling. A user with
a high $p$ is more likely to pay the upper curve rather than the
lower curve. Knowing this, he has an incentive to submit a high
probability of use and thus not to cheat.

\subsection{The reservation contract as an option}

The contract discussed in previous sections can be equivalently
regarded as an ``option''. Because $g(p)$ is the minimal amount
the user has to pay in any event, we can ask him to pay it in
period 1, and only to pay $f(p)-g(p)$ in period 2 if he needs one
unit of resource at that time. Hence, by paying an amount $g(p)$,
the user achieves the right but no the obligation to buy one unit
of resource at price $f(p)-g(p)$ in period 2. Naturally, we may
call $g(p)$ the \emph{premium} or the \emph{price of option}, and
$f(p)-g(p)$ the \emph{price of the resource}.

\begin{figure}
\psfrag{fg}{$f(p)-g(p)$}
\psfrag{g}{$g(p)$}
\centering \includegraphics[scale=.7]{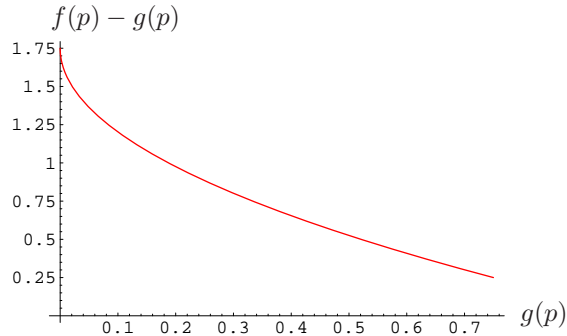}
\caption{\label{fig:curve}Price-premium curve for the user. The
horizontal axis is the premium (value of contract, value of option). The
vertical axis is the price the user pays for one unit of resource in
period 2.}
\end{figure}

Fig.~\ref{fig:curve} shows the parametric plot of resource price
versus option price, for $p\in[0,1]$. Instead of submitting an
explicit $p$, the user can equivalently choose one point on this
curve and pay accordingly. His probability $p$ can then be
inferred from his choice (using Eq.~(\ref{eq:f}) or (\ref{eq:g})).
This alternative method may be more user-friendly because people
tend to be more sensitive to monetary values rather than
probabilities. We can even further simplify the curve by providing
the user with a table with the values of a few discrete points
along the curve.

\subsection{Possible extensions}

Simple as it may seem, the two period model can  already solve a
wide range of reservation problems. Here we show how the mechanism
can be extended to solve more nontrivial problems, as when there
is uncertainty not only in the consumption of one unit, but also
in both the number of consumption units and their consumption
time.

\begin{example}
(Uncertain number of units) By checking past web statistics, a
company discovers that its website has the following pattern of
visits: 90\% of the days it needs one unit of bandwidth, 6\% of
the days two units, 3\% of the days three units, only 1\% of the
days does it need four units of bandwidth.
\end{example}

Here the company faces a four-point distribution of usage rather than a
two-point distribution (either 1 or 0) discussed in the two period
model. Imagine there are four units of resources: $U_1$, $U_2$, $U_3$
and $U_4$. The company's usage pattern can be written as $P(U_1)=0.9$,
$P(U_1U_2)=0.06$, $P(U_1U_2U_3)=0.03$, and $P(U_1U_2U_3U_4)=0.01$.
Breaking down to individual unit, the pattern is $P(U_1)=1$ (for sure
one unit will be consumed), $P(U_2)=0.1$ (with probability 0.1 the
company will need at least two units), $P(U_3)=0.04$, and $P(U_4)=0.01$.
Thus, to efficiently reserve bandwidth for some day in the future, the
company can reserve one unit for sure, and buy three options, all for
the same day, with $p=0.1$, 0.04, and 0.01 respectively.

\begin{example}
(Uncertain consumption time) A biochemist will need for sure to
use a public supercomputer to run some CPU-heavy simulations next
month, but he has no idea on which day he will need it. He wants
to reserve the supercomputer now to save cost.
\end{example}

Say that the next month has 30 days. The probability that he will
need the computer on one particular day next month is $p=1/30$. He
can buy 30 ``$p=1/30$'' options, one for each day next month. For
a numerical calculation, assume that $C>2$ and $k=2$. His total
expected cost would then be
\begin{equation}
30 \,g(\frac1{30}) + f(\frac1{30}) - g(\frac1{30})= 2- \frac 1{30}
\approx 2,
\end{equation}
so he pays less than 2 for an uncertainty over 30 days, which is
not bad.

\medskip
\noindent\underline{Remark:} The careful reader might notice that in
Example 1, the $U_i$'s are no longer independent, whereas in Example
2, the consumptions on two different days are not independent
either. This is not really a problem, because although the options
reserved by one user can be dependent, as long as the options
reserved by different users are independent, Lemma \ref{lm:conda}
still works.

\section{A Multi Period Truth-Telling Reservation}

In the previous 2-period mechanism, if a user learns more in time
about the likelihood of his needing the resource, it is impossible
for him to modify the original contract. To solve this issue we
extend our mechanism so that the user can both submit early for a
larger discount and update his probability afterwards to a more
accurate one. We thus consider a dynamic extension of the problem
in which the user is allowed to change his probability of future
use some time after his initial submission.

\subsection{The information structure}

Assume that everyone lives for three periods. In period 3 the user
might need to consume one unit of resource. He can either reserve
one unit in period 1 at price 1, or in period 2 at a price $C>1$,
or buy it in period 3 at price $C^2$.\footnote{The $(1, C, C^2)$
assumption is not essential. We could have assumed $(1, C_2, C_3)$
instead and the main result of this section will continue to hold, just
that the maths would become considerably messier.} The additional period
2 is
introduced to exploit the user's information gaining process. In
order to make this meaningful we need to carefully define the
information structure, especially what does ``gain more
information'' mean.

Suppose that each user can end up in either state $A$ or state $B$
in period 2. When in period 1, the user knows his probability of
entering each of the two states ($A$ or $B$), but he does not know
exactly which state he will enter until after period 2. In this
sense, we say that the user ``gains'' an extra bit of ``state
information'' in period 2. His probability of consuming one unit
of resource in period 3 actually depends on his state and will
thus report a more accurate probability once he gets the ``state
information''.

\begin{figure}[ht]
\centering\includegraphics[scale=.6]{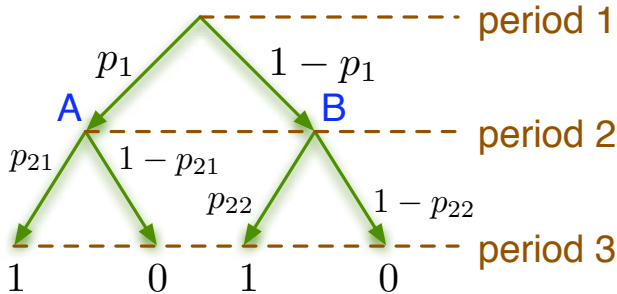}
\caption{\label{fig:info-struct-3p}The information structure for three
periods.}
\end{figure}

This information structure is depicted in
Fig.~\ref{fig:info-struct-3p}. The user enters state $A$ with
probability $p_1$ and state $B$ with probability $1-p_1$. If he
enters state $A$, with probability $p_{21}$ he will need the
resource in period 3. If he enters state $B$, he will need the
resource with probability $p_{22}$. He knows these probabilities
$(p_1, p_{21}, p_{22})$ at the beginning. Clearly, in period 1
when he has no state information, his probability of needing the
resource in period 3 is
\begin{equation}
p=p_1 p_{21} + (1-p_1) p_{22}.
\end{equation}

\begin{example}
On New Year's day, a user is struggling against a paper deadline
due on Feb 1, which is for a conference to be held on Apr 1. With
probability 0.7 he will finish the paper before the deadline, and
there is 0.6 probability that it will get accepted. If he cannot
finish it before the deadline, he can still submit a post-deadline
paper, which will only be accepted with probability 0.2. He will
be informed whether his paper is accepted some time after his
submission. He is thinking of booking a plane ticket now.
\end{example}
The three periods are Jan 1, Feb 1 and Apr 1. The probabilities are
$p_1=0.7$, $p_{21}=0.6$, and $p_{22}=0.2$.

\bigskip
\noindent \underline{Remark:} We should distinguish between the
concepts of
 \emph{error} and \emph{uncertainty}. For example, by repeatedly
 tossing an unfair coin we can estimate its probability more and more
 accurately, but even if we know its exact probability distribution we
still do not
 know what the outcome would be for the next time. That is, by
maintaining a large history we can
 reduce \emph{error} but not
 \emph{uncertainty}. In our information structure we assume that the
 user knows his accurate probabilities (no error). In this context
information is
 defined in the ``uncertainty elimination'' sense, rather than in the
``error
 elimination'' sense.

\subsection{The three period coordinator game}

Again we describe a mechanism used by a coordinator to make profit
by aggregating the user's uncertainty. A user may submit a
probability in period 1, as in the 2-period setting. Additionally,
when he enters period 2 he is allowed to update his probability
based on his new information gained at that time. This way the
user can enjoy the full discount while simultaneously  utiize
maximum information. His final payment in period 3 is determined
by the one or two probabilities he submitted. The whole mechanism
is described more rigorously as follows.

\begin{enumerate}
 \item (Period 1) The user may submit a probability $q_1$, which is
 suggested but not obligatory to be the real probability $p=p_1 p_{21} +
 (1-p_1) p_{22}$ that he will need one unit of resource in period 3.

 \item (Period 1) The coordinator reserves $q_1$ units of resource
 from the resource provider (at price 1).

 \item (Period 2) The user may submit a probability $q_2$, which is
 suggested but not obligatory to be the real probability that he will
 need one unit of resource in period 3, based on his information at
 this time (i.e. either $p_{21}$ or $p_{22}$).

 \item (Period 2) The coordinator adjusts his holdings to match the
 new probability $q_2$.

 \item (Period 3) If the user claims the need of one unit of resource,
 the coordinator delivers one reserved unit to him. If his
 reservation pool is not large enough, he buys more resource from the
provider (at
the higher unit price $C^2$) to meet the demand.

 \item (Period 3) The user pays according to Table
 \ref{tb:payment-1}:
\begin{table}[ht]
\centering
\begin{tabular}{|c|c|c|c|}
\hline
& $q_1$ not $q_2$ & $q_2$ not $q_1$ & both $q_1$ and $q_2$ \\
\hline
uses one unit & $f_1(q_1)$ & $f_2(q_2)$ & $f_1(q_1) - \alpha f_2(q_1)
+ \alpha f_2(q_2)$ \\
\hline
does not use & $g_1(q_1)$ & $g_2(q_2)$ & $g_1(q_1) - \alpha g_2(q_1) +
\alpha g_2(q_2)$ \\
\hline
\end{tabular}
\caption{\label{tb:payment-1}The user's payment table. The columns
represent his three possible submission patterns.}
\end{table}
\end{enumerate}

In Table \ref{tb:payment-1}, $(f_1, g_1)$ and $(f_2, g_2)$ are two
sets of 2-period truth-telling functions solved in
Section~\ref{sec:2period}.
\begin{eqnarray}
f_1(p) &=& 1+\frac {k_1} 2 - k_1 p + \frac{k_1 p^2}2,\\
g_1(p) &=& \frac{k_1 p^2}2,\\
f_2(p) &=& C+\frac {k_2} 2 - k_2 p + \frac{k_2
p^2}2,\\
g_2(p) &=& \frac{k_2 p^2}2,
\end{eqnarray}
where $k_1 \in [1, \min\{2(C^2-1),2\}]$ and $k_2 \in [C,
\min\{2(C^2-C),2C\}]$. To make the mathematical analysis easier,
we will choose $k_1=k \in [1, \min \{2(C-1), 2\}]$ and $k_2 = Ck
\in [C, \min\{2(C^2-C),2C\}]$, so that
\begin{eqnarray}
\label{eq:proportional-payment-1} f_2(p) &=& C f_1(p),\\
\label{eq:proportional-payment-2} g_2(p) &=& C g_1(p).
\end{eqnarray}
We require that $f_2>f_1$ and $g_2>g_1$, so that the user
pays more when he reserves late.

The third column of Table \ref{tb:payment-1} needs special notice.
The new parameter $\alpha>0$ is a ``friction parameter'' crucial
for our mechanism to work. The expression $f_1(q_1) - \alpha
f_2(q_1) + \alpha f_2(q_2)$ can be understood as follows. First,
the user signs a contract in period 1 and agrees to pay $f_1(q_1)$
if he uses one unit of resource later. Second, in period 2 the
user can adjust his probability to $q_2$ by ``selling'' some of
his old $q_1$ contracts and ``buying'' some new $q_2$ contracts.
Because he ``sells'' and ``buys'' in period 2, the selling and
buying prices should be those of period 2, namely the two $f_2$
terms. We emphasize that in the contract form solution described
here, the user only signs one contract that takes care of two
periods and does not do any trading. However, there is an
equivalent option formulation in which the user does sell his
options, which we describe in Section \ref{sec:option}.

Intuitively, because $f_2>f_1$, $\alpha$ has to be small enough since
otherwise the users would want to ``buy'' a lot of $q_1$ contracts in
period 1 and ``sell'' them later in period 2. In fact, it can be shown
that there exists a nonempty region of $\alpha$ that allows
the mechanism to work - that is, the user can save cost (even more
than using the 2-period mechanism) and the coordinator can still
profit. Formally, we have

\begin{theorem}
Suppose $\alpha \in (0, 1/C)$. The user's optimal strategy is to
submit a probability in period 1 and to adjust it in period 2. Each
probability he submits is his true probability in that period.  In
addition, the coordinator is profitable.
\end{theorem}

\begin{proof}
Follows from Lemmas \ref{lm:always-adjust},
\ref{lm:period1-truth-telling}, \ref{lm:submit-twice},
and~\ref{lm:profit} in the Appendix.
\end{proof}

\subsection{\label{sec:option}Three period options}
As for the 2-period problem, there is an equivalent ``option'' form of
the 3-period contract, which we now describe. Assume
Eq.~(\ref{eq:proportional-payment-1}) and
(\ref{eq:proportional-payment-2}).

\begin{enumerate}
 \item (Period 1) There are various options that the user can buy,
 with option price $g_1(p)$ and resource price $f_1(p)-g_1(p)$, for
 all $p\in[0,1]$. The user buys one share of $q_1$-option at price
 $g_1(q_1)$.

 \item (Period 2) The user can swap $\alpha C$ (remember $\alpha C<1$)
 share of his $q_1$-option for a $q_2$-option, by paying the difference
 price $\alpha C (g_1(q_2)-g_1(q_1))$. Then he holds  a share $(1-\alpha
C)$
 of $q_1$-options and a share $\alpha C$ of $q_2$-options.

 \item (Period 3) If the user needs one unit of resource, he executes
 his options. That is, he pays $(1-\alpha C) (f_1(q_1)-g_1(q_1))$
 using his $q_1$ option, plus $\alpha C (f_1(q_2)-g_1(q_2))$ using his
 $q_2$ option.
\end{enumerate}

It is easy to verify that this option payment plan is equivalent to
Table \ref{tb:payment-1}.

\medskip
\noindent\underline{Remarks:}

1. In period 2 when the user swaps part of his option, he does this at
no additional cost.

2. In the plan discussed above all options are issued in period
1. There can be new options issued in period 2 priced at $g_2(p)$, but
then the user should not be allowed to swap period-1 options for
period-2 options. For example, this plan does NOT work:

\begin{enumerate}
\item[2.'] (Period 2) The user can swap a fraction $\alpha$ of his
period-1 $q_1$-option for period-2 $q_2$-option, by paying the
difference price $\alpha (g_2(q_2)-g_1(q_1))$.
\end{enumerate}

\noindent The next plan does work, although a bit strange:
\begin{enumerate}
\item[2.''] (Period 2) The user can swap $\alpha C$ share of his
period-1 $q_1$-option for $\alpha$ share of period-2 $q_2$-option, by
paying the difference price $\alpha g_2(q_2)- \alpha C g_1(q_1)$.
\end{enumerate}

\subsection{Multi period options}

The option form of the 3-period contract can be easily extrapolated to
an $m$-period contract ($m>3$). Assume $\beta$ is a positive number such
that
\begin{equation}
\label{eq:beta}\beta + \cdots + \beta^{m-2} =
\frac{\beta-\beta^{m-1}}{1-\beta} < 1.
\end{equation}
Such a $\beta$ certainly exists. For example $0\le\beta\le1/2$ is enough
for Eq.~(\ref{eq:beta}) to hold for all $m$. The contract now says:

\begin{enumerate}
 \item (Period 1) There are various options that the user can buy,
 with option price $g_1(p)$ and resource price $f_1(p)-g_1(p)$, for
 all $p\in[0,1]$. The user buys one share of $q_1$-option at price
 $g_1(q_1)$.

 \item[$i$.] (Period $i$: $i=2, \dots, m-1$) The user can swap
$\beta^{i-1}$
 share of his $q_1$-option for a $q_i$-option, by paying the difference
 price $\beta^{i-1} (g_1(q_i)-g_1(q_1))$.

 \item[$m$.] (Period $m$) If the user needs one unit of resource, he
executes
 his options. That is, he pays
 \begin{equation}
 \left(1-\frac{\beta-\beta^{m-1}}{1-\beta}\right) (f_1(q_1)-g_1(q_1)) +
\sum_{i=2}^{m-1} \beta^{i-1} (f_1(q_i)-g_1(q_i)).
 \end{equation}
\end{enumerate}

\section{Mechanism Behavior}

We have seen that the truth-telling reservation mechanism helps
the user save money and the coordinator to make money, so they
both have an incentive to use it. An interesting question to ask
now is whether the resource provider himself would want to use the
reservation mechanism, playing both roles of seller and
coordinator. To answer this question we need to consider objective
functions for both the user and the seller.

\subsection{The user's utility}

In the previous sections we assumed that when it happens that the
user needs one unit of resource, he has no other choice but to buy
it. In reality if the on-spot price exceeds the user's financial
limit, he can always choose not to buy. Because of this, the
resource provider cannot set the price arbitrarily high.

Suppose the user has an expected utility in the form
\begin{equation}
u=v-c.
\end{equation}
Here, $c$ is the minimum expected price he has to pay for one unit
of resource, estimated in period 1. $v$ is the value of the unit
to him in period 1, scaled to $v\in [0,1]$ for simplicity.
Equivalently, we can use period-2 value instead of period-1 value
and write
\begin{equation}
u = v_2 p - c.
\end{equation}
If the user does not buy the resource when he needs it, his
utility is zero. We again assume that the user is risk-neutral, so
he maximizes his expected utility.

A user is completely described by his $v$ and $p$.

\subsection{The seller's problem}

\subsubsection{Direct selling} Assume that it takes the resource
provider constant cost to provide the resource, so his
profit-maximization problem becomes a revenue-maximization problem
(e.g., the cost of a flight is essentially independent of the
number of passengers on a plane). Without using the truth-telling
reservation mechanism, he can only choose a reservation price
$C_1$ and a spot price $C_2$ to maximize his revenue.

To do so he must assume a prior distribution $f(v,p)$ of the users,
where $f(v,p) \,dvdp$ is the fraction of users whose $(v,p)$ lie in
the small rectangle $(v,v+dv) \times (p,p+dp)$. Suppose that he has
complete information about the users, i.e, he knows the real
$f(v,p)$. He then faces the following maximization
problem\footnote{As in standard probability texts, here $a\wedge b$
denotes the minimum of $a$ and $b$, and $I(\cdot)$ is the indicator
function.}:
\begin{equation}
\max_{C_1, C_2} \iint dv\, dp\, f(v,p) I(v \ge C_1\wedge C_2p) C_1
\wedge C_2p.
\end{equation}
Only those users with $v \ge C_1\wedge C_2p$ will buy resources from
him, and he collects $C_1 \wedge C_2 p$ from every such user. If
$C_1>C_2$ then $C_1\wedge C_2 p = C_2p$, yielding the same revenue
as having $C_1=C_2$. Thus  without loss of generosity we can assume
$C_1\le C_2$. Also, since no one will buy the resource in period 2
if $C_2>1$, the problem can be restricted to the case $C_2\le 1$.
Hence the seller solves
\begin{equation}
\max_{0\le C_1\le C_2\le 1} \iint dv\, dp\, f(v,p) I(v \ge
C_1\wedge C_2p) C_1 \wedge C_2p.
\end{equation}

In order to carry out an explicit calculation we need to assume a
specific form for $f(v,p)$. A simple one is $f(v,p)=1$, which
implies that $v$ and $p$ are both independent and uniformly
distributed over $[0,1]$. The seller's revenue per user is thus
\begin{eqnarray}
R&=& \int_0^1 (1-C_1 \wedge C_2p)\, C_1 \wedge C_2p\, dp \nonumber \\
&=& \int_0^{C_1/C_2} (1-C_2p) C_2p \,dp + \int_{C_1/C_2}^1 (1-C_1)C_1
\,dp \nonumber \\
&=& C_1-C_1^2 + \left(\frac 23 C_1^3 - \frac 12 C_1^2 \right)
\frac 1{C_2}.
\end{eqnarray}
From this it is then not hard to check that the maximal revenue
$R_{\max}=5/24$ is achieved at $C_1=1/2$ and $C_2=1$.

\subsubsection{Options}

Within the truth-telling reservation framework, the seller sets
two prices, $f(p)$ and $g(p)$, by choosing the parameters $C_1$,
$C_2$ and $k$. Note that $C_2$ does not appear explicitly in the
prices, but only appears implicitly in the constraint $k \le
2(C_2-1)$. Thus the seller can choose a sufficiently large
$C_2$.\footnote{This may seem surprising, but remember that the
user never pays the on-spot price when he buys an option! In fact,
$C_2$ can be set greater than 1 in this case.} In the many-user
limit, his optimization problem becomes
\begin{equation}
\max_{0 \le C_1 \le 1 \le k \le 2} \iint dv\, dp\, f(v,p) I(v \ge
w(C_1, k, p)) w(C_1, k, p),
\end{equation}
where
\begin{equation}
w(C_1, k, p) = C_1 \left[ \left(1+\frac k2 \right) p - \frac k2
p^2 \right]
\end{equation}
is the expected revenue he collects from a user whose expected
value exceeds the expected cost.

Again consider the special choice $f(v,p)=1$. The seller's revenue
per person is
\begin{eqnarray}
R&=& \int_0^1 (1-w(C_1, k, p)) w(C_1, k, p) \, dp \nonumber\\
&=& \frac{C_1}2 - \frac{C_1^2}3 + \frac{C_1 k}{12} - \frac{C_1^2
k}{12} - \frac{C_1^2 k^2}{120}.
\end{eqnarray}
The related optimization problem is tedious but not hard in
principle. It is maximized at $C_1=5/8$ and $k=2$. The maximal
revenue is again $R_{\max} = 5/24$, equal to the maximum revenue
of direct selling. While this is coincidental, as we shall see in
the next section, it does show that the two revenues are
comparable.

\subsection{Other distributions}

We will now compare the two pricing schemes for other probability
distributions. Again assume that $v$ and $p$ are independent, and
$v$ is uniform on $[0,1]$. Assume now that $p$ is uniformly
distributed on $[a, b]$, where $0\le a \le b \le 1$. In other
words, assume that
\begin{equation}
f(v,p)=
\begin{cases}
\displaystyle \frac 1{b-a}, & a\le p \le b,\\
0, & \mbox{otherwise}.
\end{cases}
\end{equation}

We optimize the seller's revenue for the two schemes with multiple
choices of $a$ and $b$. The numerical results are shown in Table
\ref{tb:comparison}. It can be seen that in most cases the option
mechanism performs better than the direct mechanism. In particular
when the users' probabilities are concentrated at the small end
(row $(0,1/2)$, $(0,1/3)$ and $(0,1/5)$ in the table), the option
mechanism significantly beats direct selling. This is because in
the direct selling scheme, the seller has to compromise for a low
$C_1$ for small $p$, therefore losing considerable profit. On the
other hand, by selling options he can settle on a much higher
$C_1$ and profit from the premium.

\begin{table}[h]
\centering
\begin{tabular}{|c|c|c|}
\hline
$p$ & direct selling & options\\
\hline
$(0,1)$ & 0.208 & 0.208\\
$(0,1/2)$ & 0.167 & 0.197\\
$(1/2,1)$ & 0.250 & 0.248\\
$(0,1/3)$ & 0.130 & 0.183\\
$(1/3,2/3)$ & 0.245 & 0.246\\
$(2/3,1)$ & 0.250 & 0.250\\
$(0,1/5)$ & 0.087 & 0.141\\
$(2/5,3/5)$ & 0.248 & 0.249\\
$(4/5,1)$ & 0.250 & 0.250\\
\hline
\end{tabular}
\caption{\label{tb:comparison} The seller's revenue per person,
using direct selling or options. For example, when the users' $p$
is uniformly distributed over $(0,1/2)$, the seller's revenue per
person when using the option mechanism is 0.197.}
\end{table}

We thus conclude that the truth-telling mechanism is particularly
efficient for reservations of peak demands and rare events (small
$p$).

\section{\label{sec:concl}Conclusion}

In this paper we presented a solution to the truth revelation
problem in reservations by designing option contracts with a pricing
structure that induces users to reveal their true likelihoods that
they will purchase a given resource. Truthful revelation helps the
provider to plan ahead to reduce the risk of under-utilization and
overbooking.  In addition to its truthfulness and robustness, the
scheme can extract similar revenue to that of a monopoly provider
who has accurate information about the population's probability
distribution and uses temporal discrimination pricing.

This mechanism can be applied to any resource that exhibits bursty
usage, from IT provisioning and network bandwidth, to conference
rooms and airline and hotel reservations, and solves an
information asymmetry problem for the provider that has
traditionally led to inefficient over or under provision.

We first presented a simple two period model in which at the first
period the user knows his probability of using the resource at the
second period and purchases a reservation whose price depends on
that probability. A coordinator then aggregates the reservations
from all users and purchases the needed resources from the provider.
These resources are then delivered in the second period. In this
case, we showed how a nonlinear pricing scheme leads to both
truthful revelation on the part of the users and profitability to
both users and providers.

We then extended the two period model to a multi period model,
thus allowing for the user's likelihoods of exercising the options
to change over time. In this more realistic scheme, users are
allowed adjust their options according to updated information
about their needs while remaining truthful at each time period.

Finally we showed how this truth-telling reservation mechanism can
be interpreted in terms of standard options terminology, and concluded
that
in general it performs better than direct selling, especially for
peak-like demands.

This approach can be extended in a number of ways so as to become
useful in a number of realistic situations. With the addition of a
simulation tool developed in the context of swing
options~\cite{clearwater-05}, for example, users can anticipate
their future needs for resources at given times and price them
accordingly before committing to a reservation contract. Yet
another extension would allow for the reservation of single units
of a resource (airline seats or conference rooms, for example)
over a time interval, as opposed to a particular date.

Given the rather inefficient way through which most bursty
resources are now allocated, we believe that this mechanism will
contribute to a more useful and profitable way of allocating them
to those who need them, while giving the provider essential
information on future demand that he can then use to rationally
plan its provisioning.

\bigskip We thank Andrew Byde for valuable suggestions.

\section*{Appendix}

In this appendix we define $\delta=1/C$ to simplify the
expressions.

\begin{lemma}\label{lm:always-adjust}
If a user submits in period 1, it is weakly better for him to adjust
in period 2.
\end{lemma}
\begin{proof}
Consider a user who has arrived at period 2. He
 already submitted $q_1$ in period 1, and now can either adjust his
 probability to $q_2$, or do nothing. If he chooses to adjust, he
 will have to pay the adjustment fee in period 3, expected to be
\begin{equation}
\label{eq:adjustment-fee} p_2 \alpha [ f_2(q_2) - f_2(q_1) ] + (1-p_2)
\alpha [ g_2(q_2) - g_2(q_1) ],
\end{equation}
where $p_2= p_{21}$ or $p_{22}$ is his real probability of using the
resource in period 3, which he now knows. It can be easily checked
that, no matter what he submitted in period 1, it is always weakly
better for his to adjust $q_1$ to $p_2$ (truth-telling). Letting
$q_2=p_2$ in Eq.~(\ref{eq:adjustment-fee}), we have
\begin{eqnarray}
\nonumber & & p_2 \alpha [ f_2(p_2) - f_2(q_1) ] + (1-p_2) \alpha [
g_2(p_2) - g_2(q_1) ] \\
\nonumber &=& \alpha \{ [p_2 f_2(p_2) + (1-p_2) g_2(p_2)] - [p_2
f_2(q_1) + (1-p_2) g_2(q_1)] \} \\
&\le& 0.
\end{eqnarray}
The last step follows from Condition T1.
\end{proof}

\begin{lemma}
\label{lm:period1-truth-telling} (Period 1 truth-telling) Suppose
$\alpha<\delta$. If a user submits in period 1, he submits his real
probability.
\end{lemma}
\begin{proof}
From Lemma \ref{lm:always-adjust} we know that the user will adjust his
probability to $p_2$ in period 2. As a result his expected cost is
\begin{eqnarray}
c_{12} &=& \nonumber p f_1(q_1) + (1-p) g_1(q_1) +\\ & & \nonumber p_1
p_{21} \alpha [ f_2(p_{21}) - f_2(q_1) ] + p_1 (1-p_{21}) \alpha [
g_2(p_{21}) - g_2(q_1) ] +\\ & & \nonumber (1-p_1) p_{22} \alpha [
f_2(p_{22}) - f_2(q_1) ] + (1-p_1) (1-p_{22}) \alpha [ g_2(p_{22}) -
g_2(q_1) ]\\ &=& \nonumber p[f_1(q_1)- \alpha f_2(q_1)] + (1-p)
[g_1(q_1)- \alpha g_2(q_1)] + \mbox{function}(p_1, p_{21}, p_{22})\\
&=& \label{eq:c12-1} (1-\alpha C) [pf_1(q_1)+(1-p)g_1(q_1)] +
\mbox{function}(p_1, p_{21}, p_{22})
\end{eqnarray}

Here the notation $c_{12}$ means the user submits both in period 1
and 2. By assumption $1-\alpha C>0$. Because $(f_1, g_1)$ is
truth-telling, the last equation is minimized when $q_1=p$. Thus,
if the user submits a likelihood, he better submit $p$, the true
probability (estimated in period 1) that he will use one unit of
resource in period 3.
\end{proof}

\medskip
Note that, for the special choice $f_2=C f_1$, when the user submits
two probabilities and uses the resource, his payment can be written as
\begin{equation}
f_1(q_1) - \alpha f_2(q_1) + \alpha f_2(q_2) = (1-\alpha C) f_1(q_1) +
\alpha C f_1(q_2).
\end{equation}
Lemma \ref{lm:period1-truth-telling} assumes that $\alpha C<1$.
Then the mechanism can be understood as having the user buy
$(1-\alpha C)$ fraction of the $q_1$ contract to take advantage of
the large discount, and buys $\alpha C$ fraction of the $q_2$
contract to take advantage of his increased level of information.

\begin{lemma}
\label{lm:submit-twice} The user prefers to submit a rough estimation
 in period 1 and then adjust it in period 2, rather than to ignore
 period 1 and only submit in period 2.
\end{lemma}
\begin{proof}
We compare the user's cost in both cases. If he only
 submits in period 2, he would of course submit the true probability
 (in period 2). Thus his expected cost is
\begin{eqnarray}
\nonumber c_2 &=& p_1 w_2(p_{21}) + (1-p_1) w_2(p_{22})\\
&=& p_1 Cw_1(p_{21}) + (1-p_1) Cw_1(p_{22}).
\end{eqnarray}
If he submits in both periods, his expected payoff is
\begin{eqnarray}
\nonumber c_{12} &=& (1-\alpha C) [pf_1(p)+(1-p)g_1(p)] + \alpha c_2\\
&=& (1-\alpha C)w_1(p) + \alpha c_2,
\end{eqnarray}
where the first ``='' is obtained by using the result of Lemma
\ref{lm:period1-truth-telling} to replace $q_1$ by $p$ in
Eq.~(\ref{eq:c12-1}).

We want to show that $c_{12}<c_2$, so that the user wants to
submit twice. It can be found after some algebra that the
condition is equivalent to having
\begin{equation}\label{eq:submit-twice-1}
\frac{p_1 w_1(p_{21}) + (1-p_1) w_1(p_{22})}{w_1(p)} >
\frac{\delta-\alpha}{1-\alpha}.
\end{equation}
If $p_{21}=p_{22}$ the left hand side is 1, so the inequality is
satisfied. If $p_{21} \ne p_{22}$, then without loss of generosity we
can assume that $p_{21}<p_{22}$, and Eq.~(\ref{eq:submit-twice-1}) can
be written as
\begin{equation}
\label{eq:submit-twice-2} \frac{p_{22}-p}{p_{22}-p_{21}}
\frac{w_1(p_{21})}{w_1(p)} + \frac{p-p_{21}}{p_{22}-p_{21}}
\frac{w_1(p_{22})}{w_1(p)} > \frac{\delta-\alpha}{1-\alpha},
\end{equation}
where $p=p_1 p_{21}+(1-p_1)p_{22} \in [p_{21}, p_{22}]$. Note that
for fixed $p$, the left hand side of Eq.~(\ref{eq:submit-twice-2})
is increasing in $p_{21}$ and decreasing in $p_{22}$, so we can
let $p_{21}=0$ and $p_{22}=1$ to obtain the stronger condition
\begin{equation}
\label{eq:submit-twice-3} \frac {w_1(p)}p <
\frac{1-\alpha}{\delta-\alpha}.
\end{equation}
If Eq.~(\ref{eq:submit-twice-3}) holds for all $p$, then $c_{12}<c_2$
for all $(p_1, p_{21}, p_{22})$.

At this stage we take into account the specific form of $w_1(p)$:
\begin{equation}
w_1(p) = \left( 1+\frac k2 \right) p - \frac{k^2}2 p^2.
\end{equation}
We then have
\begin{equation}
\frac {w_1(p)}p = 1+\frac k2- \frac{k^2}2 p \le 1+\frac k2 \le C = \frac
1\delta < \frac{1-\alpha}{\delta-\alpha},
\end{equation}
where the second``$\le$'' from the fact that $k \in [1, \min
\{2(C-1), 2\}]$. Hence Eq.~(\ref{eq:submit-twice-3}) indeed holds
for all $p$, and $c_{12}<c_2$.
\end{proof}

\begin{lemma}\label{lm:profit}
The coordinator makes profit when $0<\alpha<\delta$.
\end{lemma}
\begin{proof}
The coordinator expects to collect from the user
\begin{eqnarray}
\nonumber c_{12} &=& (1-\alpha C) w_1(p) + \alpha C [p_1 w_1(p_{21}) +
(1-p_1) w_1(p_{22})]\\
\nonumber &\ge& (1-\alpha C) p + \alpha C [p_1 p_{21} + (1-p_1)
p_{22}]\\
&=& (1-\alpha C)p + \alpha Cp = p,
\end{eqnarray}
where we have used the fact $w_1(p) \ge p$ for all $p\in [0,1]$. Thus
he expects to collect $\ge p$ from each user who has probability $p$,
so he makes profit.
\end{proof}


\begin{thebibliography}{99}
\bibitem{beran-95}J. Beran, R. Sherman, M. S. Taqqu, and W. Willinger,
Long-Range Dependence in Variable-Bit-Rate Video Traffic, \emph{IEEE
Transactions on Communications}, Vol.~43, No.~2/3/4 Feb./Mar./Apr.~1995,
pp.~1566--1579 (1995).

\bibitem{conlisk-84}John Conlisk, E. Gerstner, and Joel Sobel, Cyclic
Pricing by a Durable Goods Monopolist, \emph{Quarterly Journal of
Economics} 99: 489--505 (1984).

\bibitem{clearwater-05}Scott H. Clearwater and Bernado A. Huberman,
Swing Options: A Mechanism for Pricing IT Peak Demand, \url{http://www.hpl.hp.com/research/idl/papers/swings/} (2005).

\bibitem{gale-93}Ian Gale, Advance-Purchase Discounts and Monopoly
Allocation of Capacity, \emph{American Economic Review}, Vol.~ 83(1),
pp.~135--46 (1993).

\bibitem{gribble-98}S. D. Gribble, G. S. Manku, D. S. Roselli, E. A.
Brewer, T. J. Gibson, and E. L. Miller, Self-Similarity in File Systems,
\emph{Measurement and Modeling of Computer Systems}, pp.~141--150
(1998).

\bibitem{huberman-97}B. A. Huberman and R. Lukose, Social Dilemmas and
Internet Congestion, \emph{Science}, Vol.~277, 535--537 (1997).

\bibitem{sobel-91}Joel Sobel, Durable Goods Monopoly with Entry of New
Customers, \emph{Econometrica} 59: 1455--1485 (1991).

\bibitem{varian-89}Hal Varian, Price Discrimination, Chapter 10 in R.
Schmalansee and R. Willig (eds.), \emph{The Handbook of Industrial
Organization}, Vol.~1, 597--654, Amsterdam and New York: Elsevier
Science Publishers B. V., North-Holland (1989).

\bibitem{voldman-81}J. Voldman, B. B. Mandelbrot, L. W. Hoevel, J.
Knight, and P. Rosenfeld, Fractal Nature of Software-Cache Interaction,
\emph{IBM Journal of Research and Development}, Vol.~27, No.~6,
Nov.~1981, pp.~164--170 (1981).

\bibitem{wilson-93}Robert B. Wilson, Nonlinear Pricing, pp.~377--379,
Oxford University Press (1993).
\end{thebibliography}
\end{document}